# Resource Abundance and Life Expectancy

Bahram Sanginabadi

December 2017
Economics Department
University of Hawaii at Manoa
bahram@hawaii.edu

**ABSTRACT**

This paper investigates the impacts of major natural resource discoveries since 1960 on life expectancy in the nations that they were resource poor prior to the discoveries. Previous literature explains the relation between nations wealth and life expectancy, but it has been silent about the impacts of resource discoveries on life expectancy. We attempt to fill this gap in this study. An important advantage of this study is that as the previous researchers argued resource discovery could be an exogenous variable. We use longitudinal data from 1960 to 2014 and we apply three modern empirical methods including Difference-in-Differences, Event studies, and Synthetic Control approach, to investigate the main question of the research which is "how resource discoveries affect life expectancy?". The findings show that resource discoveries in Ecuador, Yemen, Oman, and Equatorial Guinea have positive and significant impacts on life expectancy, but the effects for the European countries are mostly negative.

1. Introduction

An irony in our world is that the countries that are blessed with an easy income from natural resources are not necessarily blessed with higher welfare than the other nations without such an income whatsoever. In fact, the natural resource abundant nations are extremely diverse. Politically, these countries are in a spectrum from brutal dictatorships to free democracies. Economically, some suffer from poverty, some are rich industrial countries, and many others are in between these two extremes. This makes understanding the impacts of income from natural resources on welfare difficult to understand. According to the World Bank data, life expectancy -an element of welfare- in Norway, one of the top oil producers, in 2015 was 82.1, while in Nigeria and Angola, other top oil producers, in the same year it was only 53 and 52.6 respectively. Obviously, factors other than income from natural resources contribute to such enormous differences in life expectancy, but the question is "how natural resource discoveries affect life expectancy?". This is the main question of our study.

To the best of our knowledge no other studies attempted to answer this question before. However, two groups of studies are relevant to our study. Wile, one group of studies focus on relation between natural resource abundance and economic growth (Sachs and Warner, 1995; Velasco, 1997; Gylfason et al., 1999; Tornell and Lane, 1999; Leite and Weidmann, 1999; Ross, 2001; Papyrakis and Gerlagh, 2007; Caselli and Cunningham, 2009; Brollo et al., 2010; Vicente, 2010; Van der Ploeg, 2011; Sala-i Martin and Subramanian, 2013; Michaels, 2011; Smith, 2015) the other group try to understand the relation between income and life expectancy (Stolnitz, 1965; Demeny, 1965; Kitagawa and Philip, 1973; Cutler et al., 2006; Preston, 2007; Cristia, 2007; Mackenbach et al., 2008; Duggan, 2008; Braveman et al., 2010; Waldron, 2013; Chetty et al., 2016). These studies have essential contributions to the literature, but the main question of our research has been entirely neglected.

Nevertheless, this is a crucial question because unlike economic growth that the first mentioned group of studies above focus on and as Bloom and Canning (2006) discuss it is an "imperfect proxy for human well-being", life expectancy is a key element of welfare. life expectancy is strongly associated with well-being (Papavlassopulos and Keppler, 2011; Mackenbach and Looman, 2013). Papavlassopulos and Keppler (2011) maintain that "life is valued for its quality, and, if positive, its extension is an improvement of well-being" (p. 475). In addition, Becker et al (2005)'s finding that 'even with advent of AIDS in Africa which increased mortality, rising in life expectancy between 1960 and 2000 significantly contributed to gain in global welfare' is another support for this claim. Therefore, to understand how natural resource discoveries affect welfare it is useful to study how they impact life expectancy.

In addition, many oil-producing nations rely heavily on oil revenues. It matters to understand how being a resource abundant nation affects their welfare. This can have essential policy implications. Suppose that it turns out that in a country the impact of a major oil discovery has been negative on life expectancy. If policy makers want to improve the welfare in their society, they may think about improving the alternative sources of national income. Also, they may think about the portions of income from natural resources that they spend on health and education. The economists and policy makers are not the only groups who might care about the answer for this question, but it also could be important for people in oil-producing countries. As an example, Mahmood Ahmadinejad the former president of Iran, one of the top oil producer nations, knew that people care about the impacts of oil revenues on their well-being, chose his main presidential campaign slogan as "I will bring oil to the table of every single Iranian family".

In short, it is crucial to understand how being blessed with having an easy wealth of valuable and demanding natural resources that some seemingly lucky countries have and some unlucky ones do not, impacts welfare of people in those countries. Are the resource abundant countries truly blessed and lucky?

To answer the main question of the research, I test the following hypotheses:

A. "A major natural resource discovery has a positive impact on life expectancy"
B. "A major natural resource discovery in a poor country has a bigger impact on life expectancy than a major natural resource discovery in a rich country"

To test the mentioned hypotheses, we investigate the impacts of major natural resource discoveries since 1960 on life expectancy. To do so as precise as possible, we estimate the impacts of resource discoveries on life expectancy at birth as well as on life expectancy at birth of females and males. In addition, we estimate the impacts of resource discoveries on infant mortality, mortality under age 5, "adult, female, mortality", and "adult, male, mortality". By adding all the mentioned types of mortality, we can understand what portions of changes in life expectancy comes from mortality of each group of the population. Hence, we can get a more detailed and clear picture of the impacts of resource discoveries on life expectancy.

Whether greater income leads to higher or lower life expectancy has not be unambiguously established. Of course, Malthus is one of the earliest scholars who thought about the relation between these variables. In his Dismal Theory, Malthus, explains that there is a Negative dynamic relationship between income and mortality. In addition, at least two studies assert that there is no relationship between income per capita and mortality (Stolnitz 1965; Demeny 1965). Also, many studies have concluded that greater wealth does lead to higher life expectancy (Kitagawa and Philip 1973; Cutler et al. 2006; Preston 2007, Cristia 2007; Mackenbach et al. 2008; Duggan 2008; Braveman et al. 2010; Waldron 2013; Chetty et al 2016)

Preston[1] (2007) might be the most influential study in this area. He finds an association between national income per head and life expectancy for 1900s, 1930s, and 1960s. He has some other important findings. First, income per capita has a non-linear, positive effect on life expectancy. The effect attenuates as countries become richer. Second, the relationship has shifted upward over time. Third "Factors exogenous to a nation's level of income per head have had a major effect on mortality trends in more developed as well as in less developed countries" (p. 489). He explains that income levels per se accounts for 10 to 20 percent growth in life expectancy in the world and factors exogenous to a country's current level of income account for 75 to 90 percent. He argues that association between national income and life expectancy is indirect and if higher national income goes to public health, nutrition, education, etc. then it can decrease mortality and improve life expectancy:

> *There is no reason to expect a direct influence of national income per head on mortality; it measures simply the rate of entry of new goods and services into the household and business sectors. Its influence is indirect; a higher income implies and facilitates, though it does not necessarily entail, larger real consumption of items affecting health, such as food, housing, medical and public health services, education, leisure, health-related research and, on the negative side, automobiles, cigarettes, animal fats and physical inertia (p. 484.)*

there has been debates on the relative importance of the mentioned factors. While Preston (2007) and Deaton (2006) emphasize on the roles of public health measures, Fogel (2004) puts more weight on the of impacts of rising income on nutrition.

In addition, as we mentioned before, many studies have tried to understand the relation between natural resource abundance and life expectancy. However, whether income from natural resources leads to higher or lower life expectancy has not be clearly established. On the one hand, researchers such as Michaels (2011) and Smith (2015) find positive effects of oil discovery on economic growth. On the other hand, resource course studies argue that due to corruption and weaker institutions income from natural resources has an adverse impact on economic growth (Leite and Weidmann, 1999; Ross, 2001).

---

[1] Preston published the first version of his paper in 1975

Resource curse literature mostly grew after Sachs and Warner (1995). Some resource curse studies argue that income from natural resources can possibly alter the incentives of the leaders and make them to act in an opposite direction of well-being of their societies (Caselli and Cunningham, 2009). Also, other studies show natural resource windfalls might decrease investment and openness and have negative effects on schooling and economic growth (Papyrakis and Gerlagh, 2007). Another explanation for negative impact of natural resource discoveries on economic growth is "Dutch Disease" which argues that natural resource export tends to increase exchange rates and hence the competitiveness of industrial exports diminishes (Sachs and Warner, 1995; Gylfason et al., 1999; Van der Ploeg, 2011; Sala-i Martin and Subramanian, 2013;). In addition, some other studies maintain that a windfall of natural resources can have adverse effects on economies through political process such as increased rent-seeking (Velasco, 1997; and Tornell and Lane, 1999). Also, some other studies focus on increase in corruption and decrease in quality of the politicians because of natural resource abundance (Brollo et al. 2010; Vicente 2010). Caselli and Cunningham (2009) argue that that there could be some channels that resource rents can alter the incentives of the political leaders. Some of the channels might make the leader to invest more in the assets that could favor economic growth. However, some other channels might make the leader to invest in a way which does the opposite. Papyrakis and Gerlagh (2007) argue that abundance of natural resources increases corruption and decreases R&D expenditure, openness, schooling, and investment. Also, in more volatile economies with poor financial systems, higher corruptions, lack of rule of law, and political issues the mentioned problems could be more sever.

To summarize, the relation between natural resource abundancy and life expectancy hasn't been studied. Therefore, this is a gap in the literature. In this research, we attempt to fill this gap using advanced and new econometrics methodologies.

The main econometrics methodology that we apply is difference-in-differences fixed effect models. difference-in-differences approach estimates the impact of an event on an outcome variable by comparing the average change in the outcome variable for the treatment unit and the average change for the control group over time. We apply this approach mainly because the goal of this research is to investigate the impacts of major oil discoveries (i.e., events) on life expectancy (i.e., an outcome variable) and difference-in-differences provides the right framework to do so.

Our results show that resource discovery in Ecuador, Yemen, Oman, and Equatorial Guinea had a positive and significant effect on life expectancy. However, the impact in Denmark, Netherlands, Norway, and UK was negative.

The rest of this paper is designed as follow. Section 2 outlines the empirical design of the research. Section 3 explains the Data and treatment assignment. Section 4 provides the results and section 5 presents the conclusion of the study.

## 2. Empirical design

We apply difference-in-differences fixed effect models to answer the main question of this study. Also, because of the reasons that we will explain in this section we apply event studies and synthetic control approaches as well.

## 2.1. Difference-in-Differences

Difference-in-differences approach estimates the impact of an event on an outcome variable by comparing the average change in the outcome variable for the treatment unit and the average change for the control group over time. We apply this approach it provides a nice framework to investigate the impact of an event (resource discovery in our study) on an outcome variable (life expectancy or mortality in our study). Also, many other scholars used this approach to study the impact of an event on outcome variables. Card and Krueger (1994) might be the most famous study that used Difference-in-differences approach. They investigated the impacts of rising in Minimum Wages (an event) on Employment (an outcome variable) at Fast Food Industry in New Jersey and Pennsylvania.

Before applying these method, I divide the countries in two groups. One is the treatment and the other is control group. A treatment unit is a country which previously has been resource-poor, but after a major oil discovery becomes a resource-rich[2] nation. However, how we know how big a discovery should be to consider a country resource rich after that? Defining a threshold could be arbitrary. Therefore, I follow Smith (2015) who defines a threshold such that an oil discovery leads to producing at least 10 barrels of oil per capita. The units in the control group are some countries in the same region as the treated country, but they don't experience a major oil discovery. So, they are resource-poor countries.

Following Smith (2015) I estimate the following model to get the average effect of resource discovery on life expectancy:

$$Y_{crt} = \alpha_c + \delta Post_{ct} + \gamma_{rt} + \varepsilon_{ct} \tag{1}$$

Where $Y_{icrt}$ is the outcome variable in country c region r and year t. $\alpha_c$ is country fixed effect. $Post_{ct}$ is an indicator which takes value one if $t > T_0$ where $T_0$ is the event year (provided in table (1)) in country c. $Post_{ct}$ is zero for the years before the event. $\gamma_{rt}$ is year fixed effect dummies for region r and year t. This variable controls for the common shocks experienced across a region.

Note that for each treated country, we estimate this model seven times. We will have a different dependent (i.e., $Y_{icrt}$) variable each time. That will be the only difference between each estimation and the other ones. For the first estimation, the dependent variable is 'life expectancy at birth, total' and for the rest of the estimations we will have 'life expectancy at birth, female', 'life expectancy at birth, male', Infant mortality, Mortality under age 5, 'adult mortality, female', and 'adult mortality, female' as dependent variables respectively. For all of these separate estimations, the event, country fixed effect, and year fixed affect will be the same.

A key assumption of difference-in-differences models is common trend assumption. In fact, the identification in the difference-in-differences models relies on the common trends (parallel trends) assumption that requires the dependent variable for the treatment unit and the control group to have same trends. If the two groups have same trend, then the differences could be due to treatment.

The common trend assumption is not easy to verify. However, one can show that the outcome variable in control and treatment groups are parallel before the treatment. Even if the pretreatment trends are parallel changing policies and conditions after the treatment could affect the results. In this paper, the outcome variable which is $Y_{icrt}$ should be parallel for the treated country and control group before the treatment (i.e., is oil discovery). Regarding to the post treatment period, based on the mentioned assumption, if there was not a treatment the outcome variable should be parallel in the treated and the control units. To meet the mentioned assumption, I dropped the countries that didn't have pre-treatment parallel trends. Also, I dropped the ones that at a point their mortality increased and therefore their life expectancy decreased because of wars and diseases. For example, I dropped several African countries because of high rates of mortality from HIV/AIDS. Since the common trend assumption cannot be 100% guaranteed to hold, I use Synthetic Control Approach as well to investigate the impacts of major oil discoveries on life expectancy. A crucial advantage of this approach is that it does not require the common trend assumption to hold.

---
[2] Smith (2015) defines a threshold such that an oil discovery leads to producing at least 10 barrels of oil per capita.

Before going into synthetic control details, we explain event study approach and the reasons why we use this method.

## 2.2. Event study

We use event study method because it has two important advantages. First, it checks for pre-existing trends which could cause spurious results. Second, rather than only a post event average this method shows the temporal pattern of the treatment effect (Smith 2015).

The event study specification allows the treatment effect to vary over time:

$$Y_{crt} = \alpha_c + \delta E_{ct} + \gamma_{rt} + \varepsilon_{ct} \tag{2}$$

where $E_{ct}$ is a 3-years-period dummy variable. Here the years 1-3 before the event are omitted and the outcome variables during any other 3-years-period before and after the event will be compared with that of 1-3 years before the event.

## 2.3. Synthetic Control

We apply synthetic control approach which developed by Abadie and Gardeazabal (2003) and Abadie et al. (2015) to construct a weighted average of the countries in the control group as the synthetic group. Ideally, in this method prior to the treatment the outcome variable for the synthetic group and the treatment country overlap. After the treatment, the outcome variable of the synthetic group provides a counterfactual for the treatment country. So, by comparing these two variables the impacts of oil discovery could be understood.

As we mentioned before, applying this method we construct a weighted combination of control countries using some of their important pre-event characteristics. The initial characteristics are log of GDP per capita, percentage of the population aged 15-64, and life expectancy at birth in the years before the event. These characteristics should be chosen in a way that for the pre-event period the outcome variables of the treated country match that of the synthetic control as much as possible. To do so, we get the weights through an optimization process. For *J* control countries and *K* predictors (characteristics) the weights are found through an optimization procedure which maximizes the following function:

$$(X_1 - X_0 W)'V(X_1 - X_0 W) \tag{3}$$

where $X_1$ and $X_0$ are matrixes of the predictors in the treatment and control groups respectively. X$1$ is a *(k × 1)* matrix and X$_0$ is a *(K × J)* matrix. The goal from estimating the equation above is to get $W^*$ for the control countries. *W* which is a *(J × 1)* vector of time-invariant weights is chosen in a way that minimizes the distance between $X_1$ and $X_0W$. The weight for each country is positive and summation of all of the weights equals to one. *V* is a *(K × K)* symmetric and positive semidefinite matrix.

Given $W^*$, the treatment effect for the post-event period *t* is as follow:

$$Y_{1t} - \sum_{j=2}^{j+1} w_j^* Y_{jt} \qquad (4)$$

Where $Y_{1t}$ is the outcome variable for the treatment country in year *t*. $Y_{jt}$ is the outcome variable for country *j* in the control group in year *t*. Also, $w^*j$ is the optimized weight assigned to country *j*. I expect $Y_{1t} - \sum_{j=2}^{j+1} w_j^* Y_{jt}$ to be very close to zero before the event and for the postintervention period this difference shows the difference of the outcome variable between the treatment and the control group which represents the synthetic control estimator of the effect of the treatment. Hence, I expect them to diverge after the event in case there is a casual effect.

Synthetic control approach has at least one disadvantage. This approach could yield to biased results if the characteristics of the treated unit is far from those of the control group. To avoid this problem, we made sure that the mentioned characteristics are close to a high degree.

### 3. Data and treatment assignment

We follow Smith (2015) to define an event year. Other than that, the source of the rest of the data in this paper is the World Bank. Also, we use longitudinal data from 1960 to 2014 in all our estimations, unless stated otherwise. Smith (2015) defines a threshold for the event year such that an oil discovery leads to producing at least 10 barrels of oil per capita in a prior resource-poor nation. So, the treatment countries are the ones which have been resource-poor prior to the discovery, but they become resource-rich because of the discovery. In this study, by resource-poor we mean that the oil production of a country is less than 10 barrels per capita. Also, a resource rich nation is a country that produces more than 10 barrels of oil per capita. Even though this threshold sounds arbitrary, it is effective in separating low level producers from level ones. Smith (2015) notes that "10 barrels generate anywhere from $100 to over $800, depending on oil and gas prices in a given year. Further, most countries that pass 10 barrels per capita do so in the early stages of exploitation after a major discovery and go on to produce at much higher levels" (p. 59). we use the same event years that Smith (2015) used in his paper. Table (1) represents the initial discovery of the natural resource, the first year of production, the event years, and production and event lags.

**Table 1** – Treatment countries

|  | Event year | Initial Discovery | First production year | Production lag | Event lag |
|---|---|---|---|---|---|
| Oman | 1966 | 1963 | 1966 | 3 | 3 |
| Netherlands | 1966 | 1959 | 1963 | 4 | 7 |
| Syria | 1968 | 1959 | 1968 | 9 | 9 |
| Malaysia | 1971 | 1963 | 1970 | 7 | 8 |
| Ecuador | 1972 | 1967 | 1972 | 5 | 5 |
| Norway | 1972 | 1967 | 1971 | 4 | 5 |
| New Zealand | 1976 | 1959 | 1970 | 11 | 17 |
| United Kingdom | 1976 | 1970 | 1975 | 5 | 6 |
| Denmark | 1982 | 1966 | 1972 | 6 | 16 |
| Yemen | 1991 | 1984 | 1986 | 2 | 7 |
| Equatorial Guinea | 1992 | 1984 | 1992 | 8 | 8 |

Reference: Smith (2015)

Table (2) represents the treatment nations and the control groups associated with them. Each control group, is a set of countries in the same regions with the treatment countries. Obviously, New Zealand is not a European nation, but due to similar socioeconomic factors the European countries are the best possible nations to be considered in the control group for New Zealand.

A key advantage of our study is that it deals with natural resource discovery that plausibly is an exogenous variable. In fact, a problem associated with studying income (wealth)-life expectancy relation is endogeneity of income (wealth). Income from oil might not be random, but studies show that natural resource discovery in a resource poor country is not a factor of oil prices and economic level of nations, but to a high extend is exogenous of the countries' socioeconomic conditions (Smith, 2015).

The control group countries stay resource poor during the period of this study. Also, the countries which have been resource-rich before 1960s dropped from the sample since we don't have access to the required data long enough before they discovered natural resources. In addition, the nations that became resource-rich from other sources of natural resources are dropped from the sample.

We carefully checked the data of all the countries in the control groups and made sure that the dependent variable has parallel trends with the associated treated units. Also, we dropped the countries that mortality at a point increased because of war. For example, we dropped Cambodia and Vietnam among Eastern-Asian countries because of huge increase in mortality and decrease in life expectancy in 1970s. Also, among African countries, we dropped Rwanda due to sharp decrease in life expectancy in 1980s and 1990s and Republic of Congo because of civil wars in 1990s. In addition, we dropped Nigeria and Botswana because of significantly high death rates from HIV/AIDS.

**Table 2-** treatment and control groups

| Treated countries | Control groups |
|---|---|
| **East Asia:**<br>Malaysia | **Cambodia**, China, Hong Kong, Indonesia, Japan, Korea, Republic of, Laos, Mongolia, Philippines, Singapore, Taiwan, Thailand, Vietnam |
| **Latin America and the Caribbean:**<br>Ecuador | Costa Rica, Cuba, Dominican Republic, El Salvador, Guatemala, Honduras, Jamaica, Nicaragua, Panama, Paraguay, Puerto Rico, Uruguay. |
| **Middle East and North Africa:**<br>Yemen, Oman, Syria | Djibouti, Egypt, Israel, Jordan, Lebanon, Morocco, Tunisia, Turkey. |
| **North and Eastern Europe:**<br>Denmark, Netherlands, New Zealand, Norway, United Kingdom. | Belgium, Finland, France, Germany, Ireland, Sweden, Switzerland, Czech Republic, Hungary, Poland. Southern Europe: Greece, Italy, Portugal, Spain. |
| **Sub-Saharan Africa:**<br>Equatorial Guinea | Benin, Burkina Faso, Burundi, Cameroon, Cape Verde, Central African Republic, Chad, Cote d'Ivoire, Gambia, Ghana, Guinea, Kenya, Lesotho, Liberia, Madagascar, Malawi, Mali, Mauritania, Mauritius, Mozambique, Namibia, Niger, Senegal, Somalia, Sudan, Swaziland, Tanzania, Togo, Uganda, Zambia, Zimbabwe. |

Reference: Smith (2015)

Table (3) represents summary statistics for the treatment and control groups. Note that life expectancy at birth and infant mortality are the main dependent variables in our study. Also, we present GDP per capita and the percentage of the population that ages 15-64 because they are the main characteristics that we use to create synthetic units when we apply synthetic control approach.

**Table 3-** Summary statistics

|  | Treated Countries | | | | Control group of the treated countries | | | |
|---|---|---|---|---|---|---|---|---|
|  | GDP pc | life ex at birth | Infant mortality | Pop 15-64 | GDP pc | life ex at birth | Infant mortality | Pop 15-64 |
| Malaysia | 5355.235 | 69.060 | 24.087 | 58.591 | 6894.063 | 66.674 | 41.319 | 61.892 |
|  | (3087.907) | (4.428) | (18.229) | (5.670) | (7825.581) | (9.594) | (35.254) | (7.489) |
|  | [55] | [55] | [55] | [55] | [548] | [550] | [450] | [550] |
| Ecuador | 3761.068 | 66.171 | 58.250 | 56.633 | 4049.617 | 67.946 | 45.562 | 57.270 |
|  | (1010.499) | (7.302) | (32.613) | (4.413) | (2697.678) | (7.648) | (32.701) | (5.765) |
|  | [55] | [55] | [55] | [55] | [651] | [660] | [597] | [660] |
| Yemen | 2178.319 | 52.435 | 121.278 | 50.311 | 4394.447 | 64.437 | 60.476 | 57.225 |
|  | (692.059) | (9.676) | (72.707) | (3.442) | (3821.849) | (9.488) | (46.250) | (5.165) |
|  | [55] | [55] | [52] | [55] | [437] | [436] | [408] | [440] |
| Oman | 5616.712 | 62.918 | 66.119 | 56.941 | 4394.447 | 64.437 | 60.476 | 57.225 |
|  | (2548.245) | (10.884) | (63.769) | (7.414) | (3821.849) | (9.488) | (46.250) | (5.165) |
|  | [55] | [55] | [52] | [55] | [437] | [436] | [408] | [440] |
| Syria | 5776.062 | 66.595 | 44.949 | 52.096 | 4394.447 | 64.437 | 60.476 | 57.225 |
|  | (1530.089) | (6.618) | (31.214) | (4.422) | (3821.849) | (9.488) | (46.250) | (5.165) |
|  | [50] | [55] | [55] | [55] | [437] | [436] | [408] | [440] |
| Denmark | 17653.7 | 75.282 | 8.803 | 65.584 | 12913.2 | 74.586 | 13.852 | 65.624 |
|  | (4915.492) | (2.251) | (5.240) | (1.185) | (5893.235) | (4.066) | (12.935) | (2.494) |
|  | [55] | [55] | [55] | [55] | [740] | [770] | [741] | [770] |
| Netherlands | 16818.82 | 76.630 | 8.270 | 66.003 | 12913.2 | 74.586 | 13.852 | 65.624 |
|  | (5198.978) | (2.470) | (3.911) | (2.716) | (5893.235) | (4.066) | (12.935) | (2.494) |
|  | [55] | [55] | [55] | [55] | [740] | [770] | [741] | [770] |
| New Zealand | 14287.76 | 75.323 | 11.145 | 63.525 | 12913.2 | 74.586 | 13.852 | 65.624 |
|  | (3104.346) | (3.534) | (5.345) | (2.833) | (5893.235) | (4.066) | (12.935) | (2.494) |
|  | [55] | [55] | [55] | [55] | [740] | [770] | [741] | [770] |
| Norway | 18142.13 | 76.846 | 8.103 | 64.178 | 12913.2 | 74.586 | 13.852 | 65.624 |
|  | (7182.049) | (2.584) | (4.840) | (1.233) | (5893.235) | (4.066) | (12.935) | (2.494) |
|  | [55] | [55] | [55] | [55] | [740] | [770] | [741] | [770] |
| UK | 16138.47 | 75.410 | 10.861 | 64.660 | 12913.2 | 74.586 | 13.852 | 65.624 |
|  | (5369.084) | (3.165) | (6.081) | (1.143) | (5893.235) | (4.066) | (12.935) | (2.494) |
|  | [55] | [55] | [55] | [55] | [740] | [770] | [741] | [770] |
| E Guinea | 5584.796 | 46.801 | 107.584 | 55.340 | 1339.951 | 50.037 | 102.505 | 52.490 |
|  | (7045.76) | (6.417) | (22.864) | (2.716) | (1620.414) | (8.265) | (41.575) | (3.292) |
|  | [54] | [55] | [32] | [55] | [1574] | [1595] | [1493] | [1595] |

Notes: the numbers in the parentheses are standard deviations and the ones in the brackets are sample counts.

## 4. Results

### 4.1. Difference-in-Difference

As we already mentioned, for each treated country, we estimate equation (1) seven times. Each time we have a different dependent (i.e., $Y_{icrt}$) variable. That is the only difference between each estimation and the other ones. For the first estimation, which is represented in table 4, column (1) the dependent variable is 'life expectancy at birth, total'. For the rest of the estimations we will have 'life expectancy at birth, female', 'life expectancy at birth, male', Infant mortality, Mortality under age 5, 'adult mortality, female', and 'adult mortality, male' as dependent variables respectively. For all of these separate estimations, the event, country fixed effect, and year fixed affect are the same. Both tables 4 and 5 represent the estimations of equation (1).

The results in table 4, column (1) represent the estimation of equation (1) where the dependent variable is 'life expectancy at birth, total'. As can be seen, the results show positive post-treatment average effects of the event on life expectancy in four out of eleven treated countries in this study. these countries are Ecuador, Yemen, Oman, and Equatorial Guinea. In the post treatment period on average the life expectancies of people in the mentioned countries are higher than the no-discovery counterfactuals for around 3.4, 3.5, 3.6, and 3.2 years respectively.

Also, for five out of eleven countries in our dataset the effect of the event on 'life expectancy at birth, total' is negative. Meaning that on average their life expectancy has been lower than the no-discovery counterfactuals in the post-exploitation period. For instance, for Malaysia on average life expectancy has been about two years lower than the no-discovery counterfactual over the post-treatment period. This number is around 2.5, 2, 0.2 1.7, and 0.6 for Denmark, Netherlands, New Zealand, Norway, and UK respectively.

Columns (3) and (5) are again estimations of equation (1) where the dependent variables are 'life expectancy at birth, female' and 'life expectancy at birth, male' respectively. As can be seen, these coefficients mostly have same signs as ones in column (1). In fact, they provide more details about the post-treatment effects on life expectancy.

**Table 4-** Difference-in-Differences: Life expectancy at birth, Total, Female, Male

|  | (1) Life expectancy at birth, total | (2) $R^2$ | (3) Life expectancy at birth, female | (4) $R^2$ | (5) Life expectancy at birth, male | (6) $R^2$ | (7) $N^a$ |
|---|---|---|---|---|---|---|---|
| Malaysia | -2.292*** (0.690) | 0.957 | -1.642** (0.679) | 0.962 | -2.901*** (0.729) | 0.948 | 605 |
| Ecuador | 3.375*** (0.799) | 0.908 | 3.544*** (0.839) | 0.909 | 3.214*** (0.779) | 0.906 | 715 |
| Yemen | 3.516*** (0.743) | 0.945 | 3.171*** (0.773) | 0.943 | 3.846*** (0.719) | 0.947 | 491 |
| Arab spring [b] | 3.583*** (0.794) | 0.943 | 3.237*** (0.829) | 0.940 | 3.912*** (0.765) | 0.945 | 455 |
| Oman | 7.155*** (1.145) | 0.934 | 7.562*** (1.199) | 0.930 | 6.769*** (1.098) | 0.937 | 491 |
| Arab spring [b] | 6.844*** (1.154) | 0.931 | 7.231*** (1.213) | 0.927 | 6.474*** (1.102) | 0.935 | 455 |
| Syria | 0.579 (0.970) | 0.935 | 1.597 (0.992) | 0.936 | -0.389 (0.982) | 0.931 | 491 |
| Arab spring [b] | 1.096 (0.963) | 0.936 | 1.745 (1.008) | 0.933 | 0.478 (0.936) | 0.938 | 455 |
| Denmark | -2.554*** (0.279) | 0.943 | -2.797*** (0.252) | 0.951 | -2.324*** (0.320) | 0.931 | 825 |
| Netherlands | -2.012*** (0.419) | 0.943 | -2.077*** (0.387) | 0.948 | -1.950*** (0.474) | 0.933 | 825 |
| New Zealand | -0.285 (0.298) | 0.945 | -1.028*** (0.268) | 0.952 | 0.421 (0.343) | 0.933 | 825 |
| Norway | -1.734*** (0.325) | 0.944 | -1.976*** (0.296) | 0.951 | -1.505*** (0.371) | 0.934 | 825 |
| UK | -0.609** (0.296) | 0.945 | -1.491*** (0.267) | 0.952 | 0.230 (0.339) | 0.934 | 825 |
| E Guinea | 3.230*** (0.895) | 0.856 | 3.307*** (0.936) | 0.853 | 3.157*** (0.867) | 0.858 | 1650 |

a- Number of observations for 'life expectancy at birth, total', leb_f, and leb_m are equal.
b- Study period considered until 2010. Arab Spring refers to unrests, civil wars, and changing of political systems in several Middle Eastern and North African countries which happened after 2010. Yemen and Syria have been heavily affected by so called Arab Spring. Oman may not be affected directly, but synthetic Oman has been affected. People have been dying because of the wars. So, life expectancy might be affected. Even though, in our models I included country and year fixed effects that might capture those effects, in "Arab Spring estimations" I have estimated the models until 2010 as well.

\* Significant at 10%
\*\* Significant at 5%
\*\*\* Significant at 1%

Table 5 represents the results of estimations of equation (1) where the dependent variables are Infant mortality, Mortality under age 5, 'adult mortality, female', and 'adult mortality, male'. Column (1) shows the estimations of equation (1) where the dependent variable is infant mortality. As can be seen from the table, four out of the eleven treated countries experienced negative and statistically significant effects of the event on infant mortality. these countries are Ecuador, Yemen, Oman, and Equatorial Guinea. The equations for all mentioned countries are economically significant as well. Yemen and Oman experienced the highest levels of reduction in their infant mortalities. The results for Oman show that post-treatment infant mortality in this country was on average around 76 out of 1000 births less than the no-discovery counterfactual. Also, for Yemen, Ecuador, and Equatorial Guinea post-treatment infant mortality was on average around 53, 13, and 12 out of 1000 births lower than the non-treated counterfactuals respectively.

The results for the European countries and New Zealand shows that on average they had a higher post-treatment infant mortality than their no-discovery counterfactuals. The post-treatment infant mortality for Netherlands, New Zealand, Norway, Denmark, and UK, was around 15, 10, 12, 10, and 8 out of 1000 births less than the no-discovery counterfactuals.

In column (4) the dependent variable is mortality under age 5. Also, for the obvious reason that child mortality under age 5 is always bigger than infant mortality since the first includes the second the coefficients here are bigger than the ones in column (1). For instance, in case of Oman and Yemen, after-treatment child mortality under age 5 was on average around 111 and 83 out of 1000 births less than the non-treated counterfactuals. This number was around 10 out of 1000 births for Equatorial Guinea. Like what we saw in column (1) in case of the European countries and New Zealand, the coefficients are positive. For the European countries and New Zealand, the coefficients are just slightly bigger than the coefficients of infant mortality and they vary between 17 and 11 out of 1000 births.

Columns (7) and (10) represent the post-treatment average effect of the event on adult females and males' mortality. In most of the cases the results are in a same line with those of infant mortality and child mortality under 5 years old. Again, the biggest negative equations belong to the Middle Eastern Countries and Equatorial Guinea. Meaning that on average they had the lowest post-treatment adult mortality comparing to their non-treated counterfactuals.

For Denmark, Netherlands, and Norway, same as the results for child mortality, coefficients are small and they vary between 13 and 20 out of 1000 adults. Also, for New Zealand the coefficients are negative and the one for male mortality is statistically and economically significant. It means that on average post-discovery adult male mortality was around 25 out of 1000 less than no-discovery counterfactual.

**Table 5-** Difference-in-Differences: Infant mortality, mortality under age 5, Adult mortality, female, male

| | (1) Infant Mortality [a] | (2) R2 | (3) N | (4) Mortality under age 5 [b] | (5) R2 | (6) N | (7) Adult Mortality, female [c] | (8) R2 | (9) N | (10) Adult Mortality, male [d] | (11) R2 | (12) N |
|---|---|---|---|---|---|---|---|---|---|---|---|---|
| Malaysia | 2.107 (4.412) | 0.882 | 505 | 7.226 (7.023) | 0.863 | 505 | -6.823 (9.248) | 0.913 | 603 | 27.945** (13.946) | 0.828 | 603 |
| Ecuador | -13.681*** (3.687) | 0.897 | 652 | -24.532*** (6.224) | 0.872 | 652 | -31.563*** (7.813) | 0.874 | 715 | -14.718* (7.737) | 0.892 | 715 |
| Yemen | -53.513*** (5.293) | 0.904 | 460 | -83.526*** (8.320) | 0.899 | 460 | -32.382*** (8.948) | 0.905 | 470 | -37.399*** (8.423) | 0.920 | 470 |
| Arab spring | -50.943*** (5.588) | 0.905 | 424 | -79.136*** (8.772) | 0.901 | 424 | -33.206*** (9.560) | 0.903 | 438 | -39.136*** (8.967) | 0.918 | 438 |
| Oman | -76.234*** (9.572) | 0.893 | 460 | -111.570*** (15.289) | 0.884 | 460 | -61.937*** (13.606) | 0.882 | 470 | -52.926*** (12.817) | 0.901 | 470 |
| Arab spring | -75.596*** (9.572) | 0.895 | 424 | -110.636*** (15.301) | 0.887 | 424 | -59.242*** (13.765) | 0.879 | 438 | -49.910*** (12.903) | 0.898 | 438 |
| Syria | 7.941 (6.146) | 0.896 | 463 | 17.014* (9.982) | 0.883 | 463 | -45.890*** (11.715) | 0.887 | 470 | 21.789* (12.604) | 0.871 | 470 |
| Arab spring | 6.680 (6.104) | 0.900 | 427 | 14.911 (9.937) | 0.888 | 427 | -46.077*** (11.898) | 0.885 | 438 | 7.414 (11.370) | 0.894 | 438 |
| Denmark | 9.850*** (1.578) | 0.822 | 796 | 11.388*** (1.985) | 0.801 | 796 | 13.557*** (2.240) | 0.902 | 757 | 15.534** (6.393) | 0.816 | 757 |
| Netherlands | 15.504*** (2.376) | 0.816 | 796 | 17.671*** (2.972) | 0.797 | 796 | 16.279*** (3.256) | 0.901 | 758 | -2.057 (9.285) | 0.826 | 758 |
| New Zealand | 10.566*** (1.685) | 0.821 | 796 | 11.672*** (2.118) | 0.800 | 796 | -3.777 (2.339) | 0.906 | 759 | -25.509*** (6.730) | 0.822 | 759 |
| Norway | 12.104*** (1.838) | 0.822 | 796 | 13.773*** (2.309) | 0.801 | 796 | 16.505*** (2.526) | 0.904 | 760 | 1.125*** (7.251) | 0.825 | 760 |
| UK | 8.717*** (1.685) | 0.822 | 796 | 10.355*** (2.120) | 0.801 | 796 | -0.305 (2.300) | 0.907 | 759 | -20.914*** (6.707) | 0.822 | 759 |
| E Guinea | -11.993** (5.238) | 0.898 | 1525 | -10.889 (10.460) | 0.891 | 1525 | -59.123*** (16.622) | 0.610 | 1650 | -54.732*** (15.903) | 0.611 | 1650 |

a. Infant Mortality rate, infant (per 1,000 live births)
b. Under 5 years old Mortality rate, under-5 (per 1,000 live births)
c. Adult Female Mortality rate, adult, female (per 1,000 female adults)
d. Adult Male Mortality rate, adult, male (per 1,000 male adults)

\* Significant at 10%
\*\* Significant at 5%
\*\*\* Significant at 1%

## 4.2. Event study

Table (6) represents the results of estimation equation (2) where the dependent variable is life expectancy at birth for the entire population. This provide much more details on the effects of the event on life expectancy at birth of the treated countries.

For Malaysia, no statistical significant effect can be seen until around 40 years after the event when some significant negative coefficients show up. For Ecuador and Oman, we can see the positive and statistically and economically significant effects of the event on life expectancy. For Ecuador, until around 20 years we don't see any statistically significant effect, but for the periods after that the coefficients are all significant. For Oman, also we can see that before the event the coefficients are negative, but shortly after the event the signs change and when around 20 years from the event passed the statistically significant coefficients show up. About Yemen, as can be seen, the coefficients are statistically positive and significant, but note that the sizes of the coefficients, the significance levels, and the signs are the same even in the periods before the event. In addition, in our previous Difference-in-Differences estimations we didn't see any significant effect of the event on economic growth of Yemen. This can be because Yemen produced as much as oil just slightly less than 10 barrels per capita for a few years before the event. For Syria, except for some coefficients significant at 10 percent level, there is no other significant coefficient.

In case of European countries, Denmark, Netherlands, and Norway, however, some negative and statistically significant effects of the event can be seen. For Denmark, the significant and negative coefficients show up shortly after the event and the sizes get bigger in the next periods. For Netherlands, the sizes of the coefficients are relatively smaller than Denmark and the significant and negative coefficients don't show up until more than 30 years after the event. Meaning that for 30 years after the event life expectancy in Netherlands wasn't significantly different from no-discovery counterfactuals. For Norway, the significant coefficients show up around 15 years after the event and the absolute value of the size of the coefficients slightly get bigger as time passes. About United Kingdom, the results show no significant effect of the event on life expectancy. Note that the size and sign of the coefficients for all European countries and New Zealand match the findings from Difference-in-Differences approach in table 4. For, Equatorial Guinea, we can see some positive and statistically significant coefficients at 10 percent level around 10 years after the event.

Tables 7 and 8 provide the results of estimations of equation (2) while the dependent variables are Infant mortality and mortality under age 5 respectively. As can be seen, none of the coefficients are significant for Malaysia. Meaning that there is no evidence of significant effect of resource discovery on infant mortality in Malaysia. In addition, findings in table 8 are to a high extend similar to the ones in table 7. Ecuador, Yemen, and Oman, are the countries which got negative coefficients in table 7. Note that the findings for all countries are almost the same in table 8. Of course, because child mortality under age 5 is always bigger than infant mortality, the absolute values of the coefficients are bigger in table 8. In table 7 the coefficients for Ecuador are negative, but significant only at 10 percent level. It is the same in table 8 except that the coefficient in the last period is significant at 5 percent level. For Yemen, same as the findings in table 7, the coefficients are negative before and after the event. For Oman, however, for both case of infant mortality and child mortality under age 5 the significant coefficients show up slightly more than 10 years after the event. For Syria, again there is no significant coefficient. In both tables 7 and 8 all European countries plus New Zealand sometimes after the treatment got positive and significant coefficients. In both tables 7 and 8, For New Zealand the significant coefficients show up shortly after the event. For UK and Denmark, they show up around 3 years after the event, and for Netherlands and Norway they show up around 25 and 12 years after the event respectively. In case of Equatorial Guinea, we see no significant coefficient.

**Table 6-** Event Study, life expectancy at birth, total

| 'life expectancy at birth, total' | Malaysia | Ecuador | Yemen | Oman | Syria | Denmark | Netherlands | New Zealand | Norway | UK | E Guinea |
|---|---|---|---|---|---|---|---|---|---|---|---|
| Year-7-9 | 1.828 | -0.773 | 4.012** | -4.940** | 1.626 | -0.129 |  | -0.296 | 0.370 | 0.011 | -0.716 |
|  | (1.466) | (1.722) | (1.647) | (2.002) | (1.839) | (0.645) |  | (0.675) | (0.703) | (0.673) | (1.983) |
| Year-4-6 | 1.519 | -0.929 | 4.915*** | -3.660* | 2.028 | -0.407 | 1.580** | -0.563 | 0.321 | -0.258 | -0.514 |
|  | (1.466) | (1.722) | (1.647) | (1.999) | (1.832) | (0.645) | (0.766) | (0.675) | (0.703) | (0.673) | (1.983) |
| Year+0-2 | 0.759 | -0.962 | 5.117*** | -2.969 | 2.682 | -1.481** | 0.547 | -0.946 | -0.214 | -0.616 | -0.013 |
|  | (1.466) | (1.722) | (1.647) | (1.999) | (1.832) | (0.645) | (0.766) | (0.675) | (0.703) | (0.673) | (1.983) |
| Year+3-5 | 0.618 | -0.504 | 4.997*** | -2.023 | 3.168* | -1.960*** | 0.272 | -0.972 | -0.232 | -0.679 | 1.005 |
|  | (1.466) | (1.722) | (1.647) | (1.999) | (1.832) | (0.645) | (0.766) | (0.675) | (0.703) | (0.673) | (1.983) |
| Year+6-8 | 0.653 | 0.174 | 4.826*** | -0.743 | 3.392* | -2.327*** | -0.007 | -0.856 | -0.480 | -0.561 | 2.234 |
|  | (1.466) | (1.722) | (1.647) | (1.999) | (1.832) | (0.645) | (0.766) | (0.675) | (0.703) | (0.673) | (1.983) |
| Year+9-11 | 0.534 | 0.923 | 4.595*** | 0.644 | 3.459* | -2.458*** | 0.060 | -1.303* | -0.688 | -0.593 | 3.289* |
|  | (1.466) | (1.722) | (1.647) | (2.002) | (1.832) | (0.645) | (0.766) | (0.675) | (0.703) | (0.673) | (1.983) |
| Year+12-14 | 0.254 | 1.690 | 4.358*** | 1.789 | 3.426* | -2.966*** | 0.101 | -1.063 | -1.080 | -0.517 | 3.715* |
|  | (1.466) | (1.722) | (1.647) | (2.002) | (1.832) | (0.645) | (0.766) | (0.675) | (0.703) | (0.673) | (1.983) |
| Year+15-17 | -0.197 | 2.411 | 4.111** | 2.728 | 3.387* | -2.926*** | -0.057 | -0.237 | -1.605** | -0.326 | 3.389* |
|  | (1.466) | (1.722) | (1.647) | (1.999) | (1.832) | (0.645) | (0.766) | (0.675) | (0.703) | (0.673) | (1.983) |
| Year+18-20 | -0.394 | 3.022* | 4.066** | 3.554* | 2.946 | -3.097*** | -0.361 | -0.378 | -1.389** | -0.436 | 2.820 |
|  | (1.466) | (1.722) | (1.647) | (1.999) | (1.832) | (0.645) | (0.766) | (0.675) | (0.703) | (0.673) | (1.983) |
| Year+21-23 | -0.462 | 3.481** | 4.249** | 3.998** | 2.805 | -3.082*** | -0.512 | -0.119 | -1.463** | -0.817 |  |
|  | (1.466) | (1.722) | (1.647) | (1.999) | (1.832) | (0.645) | (0.766) | (0.675) | (0.703) | (0.673) |  |
| Year+24-26 | -0.630 | 3.780** |  | 4.707** | 2.640 | -3.177*** | -0.674 | 0.047 | -1.535** | -0.911 |  |
|  | (1.466) | (1.722) |  | (1.999) | (1.832) | (0.645) | (0.766) | (0.675) | (0.703) | (0.673) |  |
| Year+27-29 | -0.952 | 3.884** |  | 5.266*** | 2.462 | -3.022*** | -1.173 | 0.120 | -1.887*** | -0.841 |  |
|  | (1.466) | (1.722) |  | (1.999) | (1.832) | (0.645) | (0.766) | (0.675) | (0.703) | (0.673) |  |
| Year+30-32 | -1.456 | 3.146** |  | 5.712*** | 2.463 | -2.576*** | -1.454* | -0.059 | -1.456** | -1.003 |  |
|  | (1.466) | (1.517) |  | (1.999) | (1.832) | (0.645) | (0.766) | (0.675) | (0.619) | (0.673) |  |
| Year+33-35 | -2.056 | 2.697 |  | 4.998*** | 2.503 |  | -1.866** | -0.238 | -1.295* | -0.732 |  |
|  | (1.466) | (1.662) |  | (1.741) | (1.832) |  | (0.766) | (0.675) | (0.678) | (0.673) |  |
| Year+36-38 | -2.654* | 3.691** |  | 4.710** | 1.028 |  | -1.656** | -0.373 | -1.953*** | -0.900 |  |
|  | (1.466) | (1.722) |  | (1.860) | (1.832) |  | (0.667) | (0.675) | (0.703) | (0.673) |  |
| Year+39-41 | -3.113** | 3.657** |  | 6.461*** | -1.934 |  | -1.190* |  | -2.129*** |  |  |
|  | (1.466) | (1.722) |  | (1.999) | (1.832) |  | (0.713) |  | (0.703) |  |  |
| Year+42-44 |  |  |  | 6.618*** |  |  | -1.663** |  |  |  |  |
|  |  |  |  | (1.999) |  |  | (0.766) |  |  |  |  |
| Year+45-47 |  |  |  |  |  |  | -1.840** |  |  |  |  |
|  |  |  |  |  |  |  | (0.766) |  |  |  |  |
| $R^2$ | 0.9585 | 0.9107 | 0.947 | 0.942 | 0.938 | 0.944 | 0.944 | 0.945 | 0.945 | 0.945 | 0.856 |
| N | 605 | 715 | 491 | 491 | 491 | 825 | 825 | 825 | 825 | 825 | 1650 |

**Table 7-** Event Study, Infant mortality

| Infant Mortality | Malaysia | Ecuador | Yemen | Oman | Syria | Denmark | Netherlands | New Zealand | Norway | UK | E Guinea |
|---|---|---|---|---|---|---|---|---|---|---|---|
| Year-7-9 | -4.261 | 5.351 | -41.803*** | | | 2.354 | | 1.608 | -3.508 | 1.839 | 18.304* |
| | (9.393) | (7.997) | (11.195) | | | (3.652) | | (3.829) | (3.993) | (3.829) | (10.055) |
| Year-4-6 | -4.256 | 2.912 | -49.736*** | | -8.287 | 4.085 | -10.036** | 4.082 | -1.138 | 4.213 | 15.151 |
| | (9.393) | (7.981) | (11.195) | | (11.674) | (3.652) | (4.352) | (3.827) | (3.990) | (3.826) | (10.055) |
| Year+0-2 | -4.041 | 0.809 | -55.520*** | 13.281 | -12.476 | 7.751 | -3.233 | 8.311** | 2.609 | 7.475* | 11.623 |
| | (9.324) | (7.981) | (11.195) | (13.067) | (11.674) | (3.645) | (4.349) | (3.827) | (3.984) | (3.826) | (10.055) |
| Year+3-5 | -3.151 | -1.046 | -55.640*** | -3.869 | -14.760 | 9.197** | -0.886 | 10.108*** | 4.294 | 8.438** | 7.782 |
| | (9.324) | (7.981) | (11.195) | (13.067) | (11.674) | (3.645) | (4.344) | (3.824) | (3.984) | (3.823) | (10.055) |
| Year+6-8 | -8.538 | -2.052 | -56.553*** | -16.942 | -9.765 | 10.154*** | 1.214 | 11.144*** | 6.148 | 8.708** | 4.727 |
| | (9.264) | (7.981) | (11.195) | (13.039) | (11.571) | (3.645) | (4.344) | (3.820) | (3.984) | (3.819) | (10.055) |
| Year+9-11 | -10.917 | -4.024 | -60.690*** | -25.678** | -7.830 | 10.378*** | 3.399 | 11.690*** | 7.941** | 9.220** | 2.935 |
| | (9.236) | (7.981) | (11.195) | (12.946) | (11.532) | (3.645) | (4.344) | (3.820) | (3.978) | (3.819) | (10.055) |
| Year+12-14 | -10.469 | -6.561 | -65.807*** | -37.108*** | -5.497 | 10.742*** | 5.620 | 11.747*** | 9.694** | 9.444** | 2.474 |
| | (9.236) | (7.981) | (11.195) | (12.928) | (11.532) | (3.645) | (4.344) | (3.820) | (3.978) | (3.819) | (10.055) |
| Year+15-17 | -9.343 | -8.367 | -70.461*** | -47.758*** | -2.688 | 11.290*** | 7.213* | 11.637*** | 10.656*** | 9.435** | 1.942 |
| | (9.236) | (7.981) | (11.195) | (12.928) | (11.532) | (3.645) | (4.338) | (3.820) | (3.978) | (3.819) | (10.055) |
| Year+18-20 | -7.343 | -10.055 | -74.715*** | -55.621*** | 0.166 | 11.754*** | 8.199* | 12.035*** | 10.513*** | 9.999*** | 1.039 |
| | (9.236) | (7.981) | (11.195) | (12.928) | (11.532) | (3.645) | (4.338) | (3.820) | (3.978) | (3.819) | (10.055) |
| Year+21-23 | -4.980 | -12.009 | -78.161*** | -60.746*** | 2.107 | 12.132*** | 8.927** | 12.383*** | 10.501*** | 10.780*** | |
| | (9.236) | (7.981) | (11.195) | (12.928) | (11.532) | (3.645) | (4.338) | (3.820) | (3.978) | (3.819) | |
| Year+24-26 | -2.358 | -13.861* | | -64.258*** | 3.453 | 12.154*** | 9.551** | 12.680*** | 11.070*** | 11.211*** | |
| | (9.236) | (7.981) | | (12.928) | (11.532) | (3.645) | (4.338) | (3.820) | (3.978) | (3.819) | |
| Year+27-29 | -0.410 | -14.530* | | -66.279*** | 4.870 | 12.144*** | 10.339** | 12.959*** | 11.625*** | 11.656*** | |
| | (9.236) | (7.981) | | (12.928) | (11.532) | (3.645) | (4.338) | (3.820) | (3.978) | (3.819) | |
| Year+30-32 | 1.401 | -12.119* | | -66.483*** | 6.345 | 12.156*** | 11.108** | 13.247*** | 9.706*** | 11.844*** | |
| | (9.236) | (7.029) | | (12.928) | (11.532) | (3.645) | (4.338) | (3.820) | (3.978) | (3.819) | |
| Year+33-35 | 4.064 | -10.857 | | -65.721*** | 7.820 | | 11.697*** | 13.437*** | 8.827** | 11.768*** | |
| | (9.236) | (7.698) | | (12.928) | (11.532) | | (4.338) | (3.820) | (3.503) | (3.819) | |
| Year+36-38 | 6.920 | -15.109* | | -52.812*** | 9.182 | | 9.837*** | 13.449*** | 12.125*** | 11.513*** | |
| | (9.236) | (7.981) | | (12.256) | (11.532) | | (4.778) | (3.820) | (3.978) | (3.819) | |
| Year+39-41 | 9.120 | -15.485* | | -44.646*** | 10.349 | | 8.922** | | 12.058*** | | |
| | (9.236) | (7.981) | | (12.017) | (11.532) | | (4.038) | | (3.978) | | |
| Year+42-44 | | | | -59.917*** | 11.541 | | 12.130*** | | | | |
| | | | | (12.928) | (11.532) | | (4.338) | | | | |
| Year+45-47 | | | | -57.592*** | | | 12.063*** | | | | |
| | | | | (12.928) | | | (4.338) | | | | |
| R2 | 0.885 | 0.898 | 0.913 | 0.906 | 0.899 | 0.823 | 0.822 | 0.822 | 0.822 | 0.823 | 0.898 |
| N | 505 | 625 | 460 | 460 | 463 | 796 | 796 | 796 | 796 | 796 | 1525 |

**Table 8-** Event Study, Mortality under age 5

| Mortality under 5 | Malaysia | Ecuador | Yemen | Oman | Syria | Denmark | Netherlands | New Zealand | Norway | UK | E Guinea |
|---|---|---|---|---|---|---|---|---|---|---|---|
| Year-7-9 | -8.204 | 9.352 | -58.770*** | | | 3.239 | | 2.155 | -3.955 | 2.505 | 12.040 |
| | (14.918) | (13.505) | (17.696) | | | (4.595) | | (4.817) | (5.018) | (4.818) | (20.084) |
| Year-4-6 | -8.237 | 5.643 | -72.374*** | | -14.704 | 5.129 | -11.715** | 5.114 | -0.891 | 5.497 | 8.699 |
| | (14.918) | (13.479) | (17.696) | | (18.927) | (4.595) | (5.467) | (4.813) | (5.015) | (4.814) | (20.084) |
| Year+0-2 | -7.496 | 1.316 | -82.603*** | 21.124 | -21.060 | 9.261** | -3.167 | 9.811** | 3.605 | 9.294* | 4.585 |
| | (14.809) | (13.479) | (17.696) | (20.018) | (18.927) | (4.586) | (5.463) | (4.813) | (5.008) | (4.814) | (20.084) |
| Year+3-5 | -5.853 | -2.327 | -83.361*** | -3.415 | -24.021 | 10.954** | -0.459 | 11.756** | 5.323 | 10.405** | -0.412 |
| | (14.809) | (13.479) | (17.696) | (21.018) | (18.927) | (4.586) | (5.457) | (4.810) | (5.008) | (4.811) | (20.084) |
| Year+6-8 | -14.099 | -4.939 | -85.553*** | -23.157 | -14.501 | 12.015*** | 1.995 | 12.811*** | 7.382 | 10.661** | -4.712 |
| | (14.713) | (13.479) | (17.696) | (20.973) | (18.760) | (4.586) | (5.457) | (4.805) | (5.008) | (4.806) | (20.084) |
| Year+9-11 | -17.012 | -8.706 | -92.961*** | -36.644* | -10.847 | 12.204*** | 4.280 | 13.270*** | 9.401* | 11.154** | -6.616 |
| | (14.669) | (13.479) | (17.696) | (20.824) | (18.696) | (4.586) | (5.457) | (4.805) | (4.999) | (4.806) | (20.084) |
| Year+12-14 | -15.353 | -13.060 | -101.90*** | -54.769*** | -6.530 | 12.525*** | 6.705 | 13.199*** | 11.360** | 11.415** | -5.490 |
| | (14.669) | (13.479) | (17.696) | (20.794) | (18.696) | (4.586) | (5.457) | (4.805) | (4.999) | (4.806) | (20.084) |
| Year+15-17 | -12.604 | -16.190 | -109.986*** | -71.240*** | -1.413 | 13.115*** | 8.424 | 12.987*** | 12.436** | 11.337** | -3.538 |
| | (14.669) | (13.479) | (17.696) | (20.794) | (18.696) | (4.586) | (5.449) | (4.805) | (4.999) | (4.806) | (20.084) |
| Year+18-20 | -8.671 | -18.800 | -116.986*** | -83.607*** | 3.512 | 13.684*** | 9.417* | 13.375*** | 12.115** | 11.992** | -1.940 |
| | (14.669) | (13.479) | (17.696) | (20.794) | (18.696) | (4.586) | (5.449) | (4.805) | (4.999) | (4.806) | (20.084) |
| Year+21-23 | -4.267 | -21.697 | -122.378*** | -90.957*** | 6.895 | 14.087*** | 10.160* | 13.765*** | 12.008** | 12.882*** | |
| | (14.669) | (13.479) | (17.696) | (20.794) | (18.696) | (4.586) | (5.449) | (4.805) | (4.999) | (4.806) | |
| Year+24-26 | 0.166 | -24.800* | | -95.290*** | 9.337 | 14.113*** | 10.739** | 14.034*** | 12.577** | 13.418*** | |
| | (14.669) | (13.479) | | (20.794) | (18.696) | (4.586) | (5.449) | (4.805) | (4.999) | (4.806) | |
| Year+27-29 | 3.810 | -25.333* | | -97.223*** | 11.732 | 14.096*** | 11.565** | 14.370*** | 13.201*** | 13.920*** | |
| | (14.669) | (13.479) | | (20.794) | (18.696) | (4.586) | (5.449) | (4.805) | (4.999) | (4.806) | |
| Year+30-32 | 6.951 | -20.999* | | -96.778*** | 14.162 | 14.046*** | 12.434** | 14.696*** | 11.052** | 14.180*** | |
| | (14.669) | (13.471) | | (20.794) | (18.696) | (4.586) | (5.449) | (4.805) | (4.403) | (4.806) | |
| Year+33-35 | 10.814 | -18.767 | | -95.153*** | 16.399 | | 13.091** | 14.913*** | 10.038** | 14.063*** | |
| | (14.669) | (13.999) | | (20.794) | (18.696) | | (5.449) | (4.805) | (4.821) | (4.806) | |
| Year+36-38 | 14.877 | -26.009* | | -76.252*** | 18.370 | | 11.008** | 14.896*** | 13.727*** | 13.746*** | |
| | (14.669) | (13.479) | | (18.106) | (18.696) | | (5.449) | (4.805) | (4.999) | (4.806) | |
| Year+39-41 | 18.014 | -26.469** | | -64.619*** | 19.991 | | 9.943** | | 13.660*** | | |
| | (14.669) | (13.479) | | (19.329) | (18.696) | | (5.072) | | (4.999) | | |
| Year+42-44 | | | | -86.990*** | 22.095 | | 13.517** | | | | |
| | | | | (20.794) | (18.696) | | (5.449) | | | | |
| Year+45-47 | | | | -84.098*** | | | 13.417** | | | | |
| | | | | (20.794) | | | (5.449) | | | | |
| R² | 0.867 | 0.873 | 0.907 | 0.896 | 0.887 | 0.802 | 0.801 | 0.801 | 0.801 | 0.802 | 0.891 |
| N | 505 | 652 | 460 | 460 | 463 | 796 | 796 | 796 | 796 | 796 | 1525 |

### 4.3. Synthetic Control

Figure 1 represents the main findings of synthetic control approach. Here we have provided the results of this approach for 'life expectancy at birth, total'. The findings of this approach for 'life expectancy at birth, female' and 'life expectancy at birth, male' can be find in Appendix 1. In Figure 1, the vertical dashed line shows the same event years that we represented in table 1. For each country, the solid curve shows life expectancy for that country, while the dashed curve represents the synthetic unit. For the years after the event, the difference between the solid and the dashed curves represents the outcome of interest in each year. In case of Malaysia, for couple of years after the event 'life expectancy at birth, total' has been higher than the synthetic unit, but in the last years of the period it has been smaller. In Ecuador, however, for all years after the event 'life expectancy at birth, total' has been higher than the synthetic Ecuador. Also, note that the difference between the curves has been increasing until the last year of the study.

For Yemen, Synthetic control approach doesn't represent a good pre-treatment overlap. Therefore, the post-treatment difference between the curves cannot be interpreted as a positive effect of the event on life expectancy. In case of Oman, for several years 'life expectancy at birth, total' has been lower than the synthetic unit, but it passes the counterfactual in the last years of the study. 'life expectancy at birth, total' for Syria and the Synthetic unit overlap before the event, but they don't strongly diverge after the event. This perhaps implies that oil discovery didn't have any significant effect on life expectancy of people in Syria.

The findings of Synthetic Control approach show that among European countries Denmark, Norway, and Netherlands had a lower post-treatment life expectancy rather than their synthetic unites. For UK, however, the curves overlap and diverge here and there. Also, for New Zealand, the post-treatment 'life expectancy at birth, total' has been lower than the Synthetic unit, but later this changes in the last years of our study. The post treatment 'life expectancy at birth, total' for Equatorial Guinea has been bigger than that of synthetic unit for more than a decade. This also changes in the last years of the study.

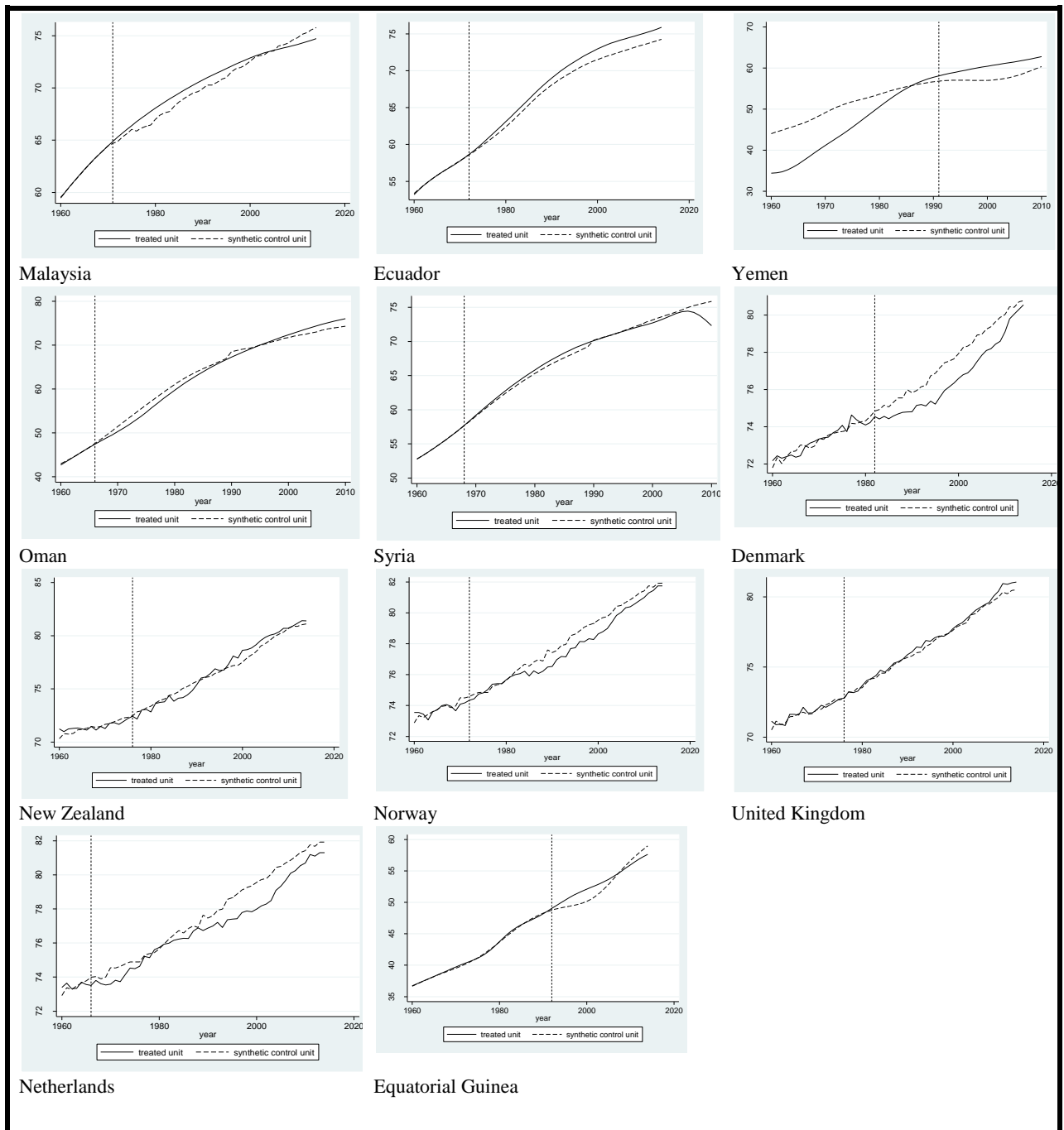

Fig 1- Synthetic control's main results. The variable on vertical axis is life expectancy at birth for the entire population ('life expectancy at birth, total') and horizontal axis represents year. The straight vertical dashed line shows the event year. The solid curve shows 'life expectancy at birth, total', while the dashed graph represents the counterfactual (synthetic unit). Note that these graphs do not provide any information about significance of the effect of the event.

## 4.4. Obtaining stationary residuals

Table 9 represents unit root tests results. The reason behind providing this table is that the sample in this paper includes many time periods and if the residuals of the regressions are not stationary then the findings might not be consistent. So, I do Levin-Lin-Chu Unit Root test on residuals of the Difference-in-Differences regressions from the estimations of equation (1). The null hypothesis here is that "Panels contain unit roots" and the alternative hypothesis is "Panels are stationary". The first column in table 9 shows the results of the unit root test for 'life expectancy at birth, total'. As can be seen, for most of the countries the residuals are stationery since we can reject the null hypothesis at 1% significance level, but for the European countries the residuals are integrated of a degree more than zero which means they are not stationary. In other words, we cannot reject the null hypothesis that "Panels contain unit roots". The second column, shows the results of the unit root test for infant mortality. As can be seen, for all the countries the residuals are stationary. In addition, the last column provides the findings of the unit root test for child mortality under age 5. Here, the residuals are stationary for all the countries except Equatorial Guinea. The findings from this table show that estimating the effects of resource discovery on infant mortality is very crucial in this study. Because the findings from these estimations could be consistent and therefore they provide a strong robustness tool for estimating the impacts of resource discovery on life expectancy.

**Table 9**- Levin-Lin-Chu Unit Root test: Adjusted t*

|  | Life expectancy, total | Infant Mortality | Mortality under age 5 |
|---|---|---|---|
| Malaysia | -38.115*** | -33.100*** | -31.825*** |
|  | 0.0000 | 0.0000 | 0.0000 |
| Ecuador | -55.568*** | -36.163*** | -53.254*** |
|  | 0.0000 | 0.0000 | 0.0000 |
| Yemen | -30.306*** | -17.414*** | -20.325*** |
|  | 0.0000 | 0.0000 | 0.0000 |
| Oman | -29.898*** | -26.053*** | -28.476*** |
|  | 0.0000 | 0.0000 | 0.0000 |
| Syria | -35.586*** | -50.521*** | -49.217*** |
|  | 0.0000 | 0.0000 | 0.0000 |
| Denmark | 28.603 | -27.651*** | -29.814*** |
|  | 1.0000 | 0.0000 | 0.0000 |
| Netherlands | 24.198 | -49.132*** | -46.185*** |
|  | 1.0000 | 0.0000 | 0.0000 |
| New Zealand | 28.453 | -24.067*** | -24.604*** |
|  | 1.0000 | 0.0000 | 0.0000 |
| Norway | 24.590 | -31.552*** | -31.755*** |
|  | 1.0000 | 0.0000 | 0.0000 |
| UK | 25.219 | -19.116*** | -21.252*** |
|  | 1.0000 | 0.0000 | 0.0000 |
| E Guinea | -2.454*** | -3.801*** | 0.5230 |
|  | 0.0071 | 0.0001 | 0.6995 |

H0: Panels contain unit roots. Ha: Panels are stationary
***: reject the null hypothesis at 1% significance level
**: reject the null hypothesis at 5% significance level
*: reject the null hypothesis at 10% significance level

## 5. Conclusion

This paper uses sophisticated and modern empirical methods to investigate the impacts of oil discovery on life expectancy. The previous studies remained silent about the main question of this research which is "how natural resource discoveries affect life expectancy". Some researchers such as Preston (2007) and Deaton (2006) discussed the channels that national income affect life expectancy. In this study, we test the following hypothesis:

A. "A major natural resource discovery has a positive impact on life expectancy"
B. "A major natural resource discovery in a poor country has a bigger impact on life expectancy than a major natural resource discovery in a rich country"

We use longitudinal data from 1960 to 2014 and we apply three modern empirical methods including Difference-in-Differences, Event studies, and Synthetic Control approach to test the mentioned hypotheses.

Here we briefly explain the main findings for each country:

**Malaysia:**

The results show a negative effect of the event on life expectancy in Malaysia. Based on the results from Difference-in-Differences estimations, the post treatment life expectancy of Malaysians is on average 2.2 years lower than no-discovery counterfactual. However, even though, the coefficients of infant mortality and mortality under age 5 are positive they are not significant. Hence lower life expectancy of Malaysians cannot be explained by child mortality. Nevertheless, adult male mortality can explain the difference since after-treatment males' mortality is around 30 out of 1000 more than the no-discovery counterfactual.

**Ecuador**

The results show a positive effect of oil exploitation on life expectancy in Ecuador. Post-treatment life expectancy of Ecuadorians on average has been around 3.3 years higher than no-discovery counterfactual. Higher post-treatment infant mortality, and 'adult mortality, female' can easily explain this.

**Yemen**

For Yemen, our Difference-in-Differences results show that post-treatment life expectancy on average has been around 3.5 years higher than the no-discovery counterfactual. Also, the results from investigating the effects of event on child mortality confirms this finding since infant mortality has been around 53 out of 1000 births less than the counterfactual. In addition, under age 5 child mortality is around 83 out of 1000 births less than the no-discovery counterfactual. However, findings from Event studies show that even before the event the coefficient for life expectancy is positive and significant. Also, for child mortality the coefficients before the event are negative and significant. A look at Yemen's oil production makes it clear that oil production in 1988 heavily improved and during the years 88, 89, and 90 it was just slightly less than 1991's production which we consider it as the event year. We estimated the regressions considering year 1988 as the event year. In this new regression for the second period before the event the coefficient is not significant, but even though the size of the coefficient is smaller it is still significant at 5% level for the period 4-7 years before the event.

**Oman**

The results show strong post-treatment effect of the event on life expectancy in Oman. The Difference-in-Differences estimations show that post-treatment life expectancy is on average around 7 years higher than the non-treated counterfactual. Also, Child mortality regressions confirm this finding. The event studies show that significant decrease in child mortality shows up around 10 years after the event and significant increase in life expectancy happens around 20 years after the event.

**Syria**

For Syria, the results do not show significant effects of the treatment on life expectancy and child mortality.

**Denmark**

For Denmark, the results show negative effect of the event on life expectancy. According to The Difference-in-Differences estimations after-treatment life expectancy in Denmark on average has been 2.5 years lower than the non-treated counterfactual. In Denmark after-treatment child mortality and adult mortality on average have been respectively around 10 out of 1000 births and 15 out of 1000 adults more than no-discovery counterfactual.

**Netherlands**

Almost same as Denmark after-treatment life expectancy has been on average around 2 years lower than the non-treated counterfactual. Also, the findings show that positive mortality coefficients can explain this finding.

**New Zealand**

For New Zealand, the effect of the event on life expectancy is not significant. Post-treatment child mortality is around 10 out of 1000 births more than the no-discovery counterfactual. This can be explained by mail adult mortality since post-treatment adult male mortality has been around 25 out of 1000 adults less than the non-treated counterfactual.

**Norway**

For Norway, the results show negative effect of the event on life expectancy. According to the Difference-in-Differences estimations after-treatment life expectancy in Norway on average has been 1.7 years lower than the non-treated counterfactual. In Norway after-treatment child mortality and adult female mortality on average have been respectively around 12 out of 1000 births and 16 out of 1000 adults more than no-discovery counterfactual. Adult male mortality has been only 1 out of 1000 adults more than the counterfactual. A coefficient that even though statistically is significant, economically is not too significant.

**United Kingdom**

The effect of event on life expectancy in United Kingdom is negative too. However, the coefficient is small and it shows that after-treatment life expectancy was on average around 0.6 year less than non-treated counterfactual.

**Equatorial Guinea**

The results for Equatorial Guinea show positive effect of the event on life expectancy. According to the Difference-in-Differences estimations the after-treatment life expectancy in Equatorial Guinea has been on average around 3.2 years higher than the non-treated counterfactual. The findings from Difference-in-Differences estimations for mortality explain this. Infant mortality, and adult female and male mortalities have been respectively on average around 12, 59, and 54 out of 1000 each less than no-discovery counterfactual.

All in all, the results show that resource discoveries in Ecuador, Yemen, Oman, and Equatorial Guinea had positive and significant impacts on life expectancy, but the effect in all four European countries including Denmark, Netherlands, Norway, and UK was negative. Maybe it sounds hard to believe that impact in European countries has been negative, but note that these results have been driven from comparing these countries and their counterfactuals which are rich industrial economies, but they didn't become resource rich at a point in their history. Therefore, our first hypothesis cannot get rejected for Ecuador, Yemen, Oman, and Equatorial Guinea, but it gets rejected for the rest of the treated countries. The second hypothesis is more controversial since the impacts for some of developing Middle Eastern countries were positive, but for rich European countries were negative.

An obvious general lesson from the results is that most of the countries that were not developed prior to the resource discovery experienced a significant increase in their life expectancy, but the results show that the nations that they were rich prior to oil discovery could enjoy a higher life expectancy if they never discovered oil. A crucial point that can be seen in Fig 1 and Appendix 1 is that these countries are catching up with their counterfactuals.

An explanation for why these countries experience a lower life expectancy is that it could be partially relevant to volatility of oil price. The OECD data shows that after oil discoveries the mentioned countries experienced higher consumer prices. Also, the prices were more volatile than the other years. More expensive food, health, etc. can decrease the demand for these products and decrease health, increase mortality, and therefore decrease life expectancy.

About the developing countries they were poor prior to oil discovery and they gained welfare after the treatment. Most likely it is because they didn't have a good health system, clean water, good nutrition, education etc. prior to the event. However, the income from oil helped them to improve these systems and enhance their life expectancies.

**Appendix 1**

Extra graphs from synthetic control approach:

For each country, the graph in the left is the gap between life expectancy of the treated country and the synthetic unit. The one at the middle shows female life expectancy of the treated country in solid curve and the synthetic unit in dashed curve. The graph at the left shows male life expectancy in solid curve and the synthetic unit in dashed curve.

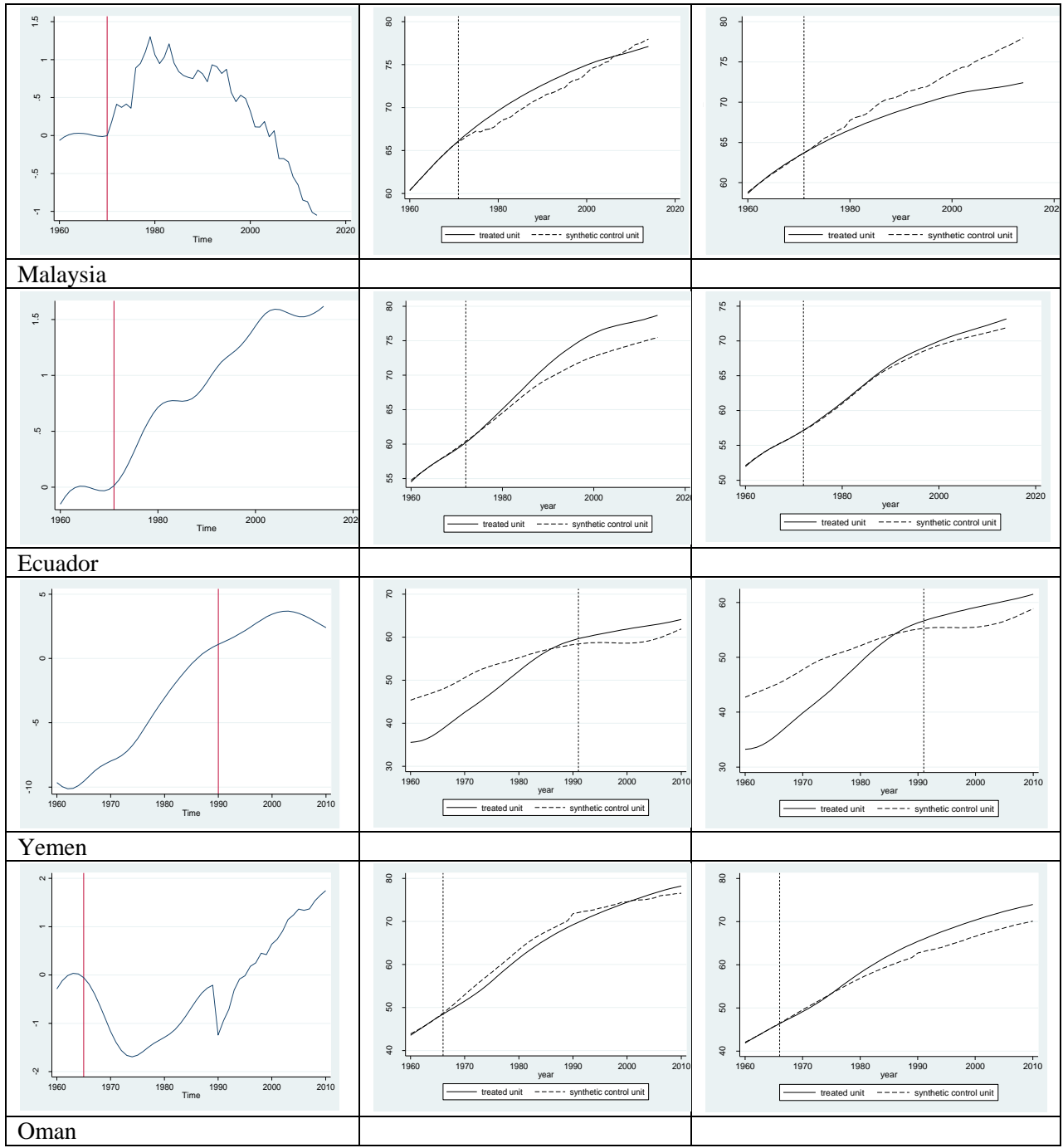

Malaysia

Ecuador

Yemen

Oman

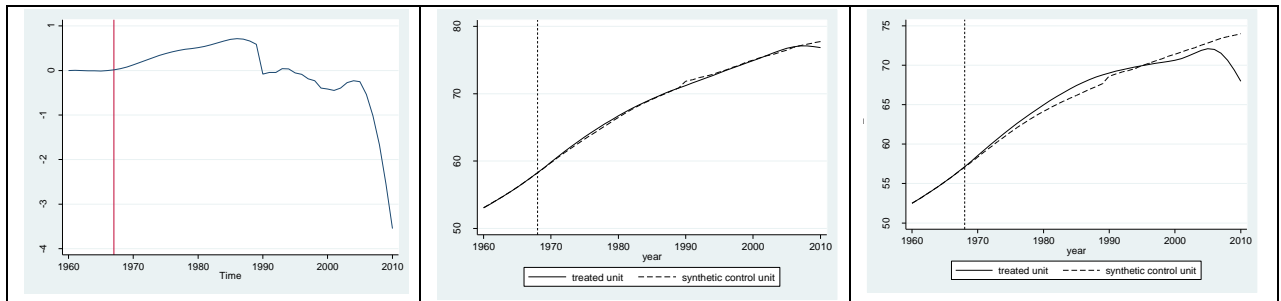
Syria

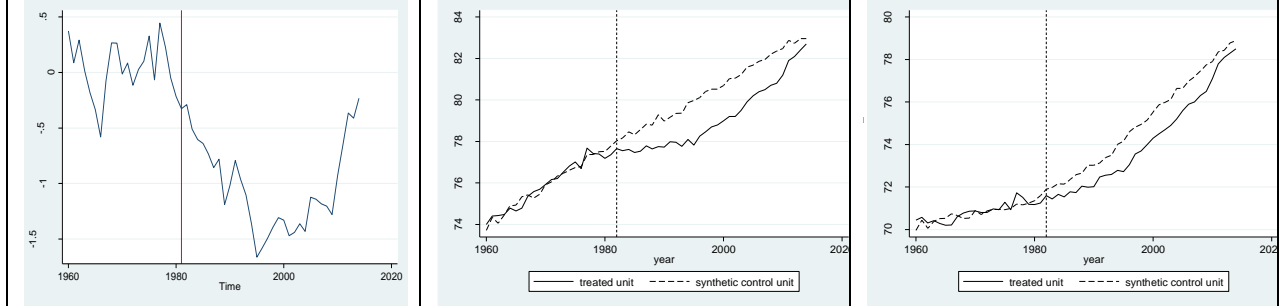
Denmark

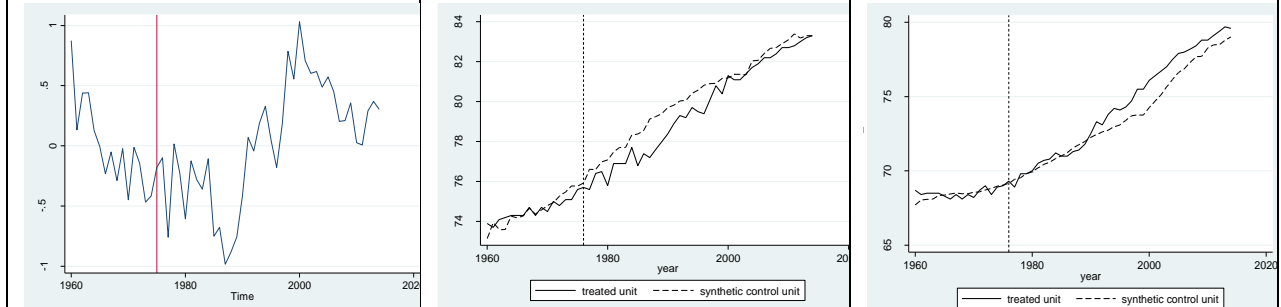
New Zealand

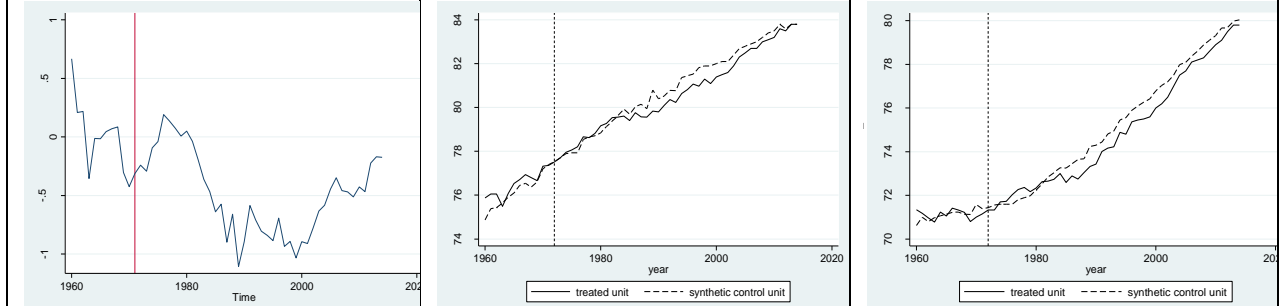
Norway

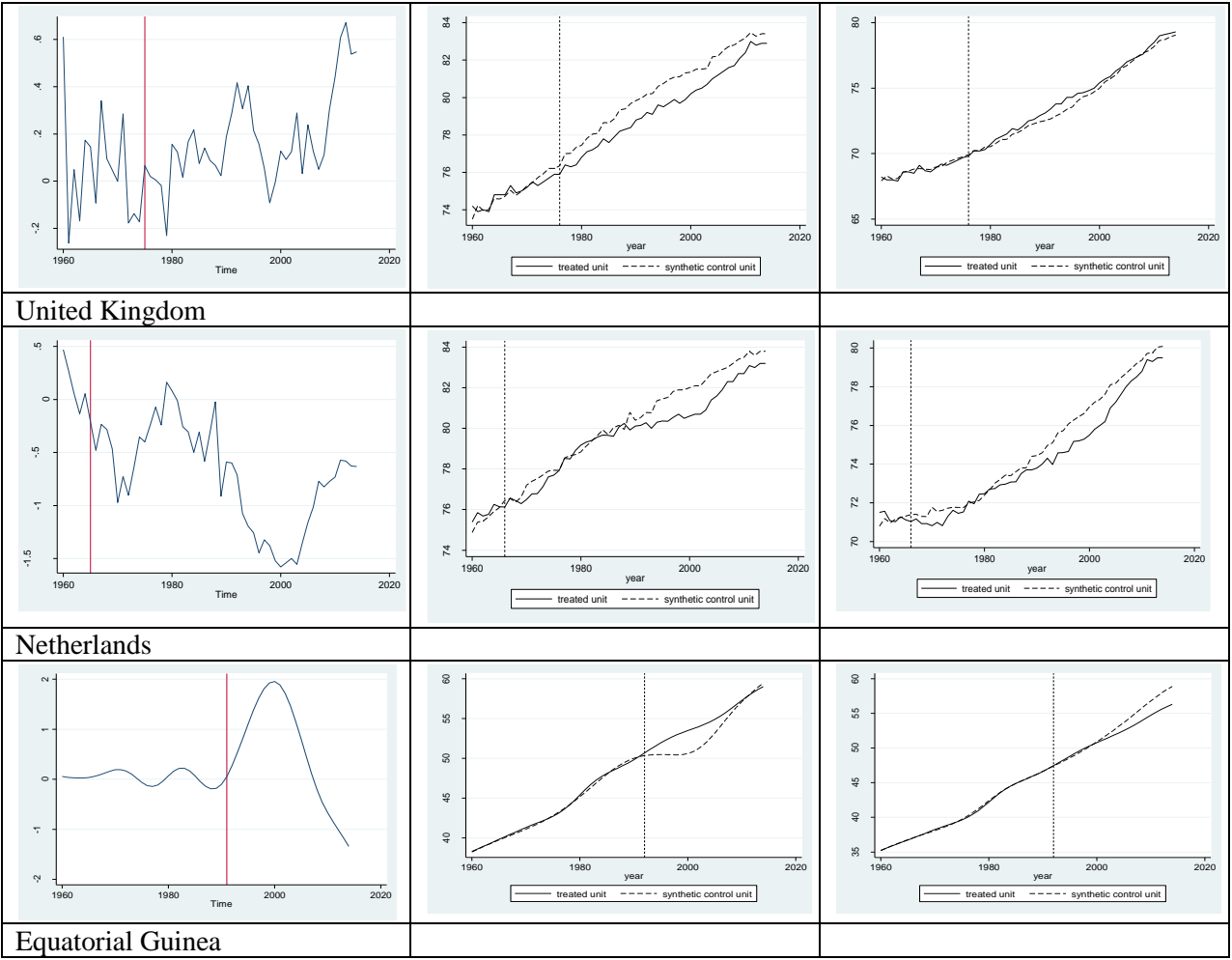

**Acknowledgment**

The author thanks Andrew Mason, Sang-Hyop Lee, Inessa Love, Nori Tarui, and Ekaterina Sherstyuk for valuable inputs and comments. The author, also, thanks the Education and Research programs' staff at the East West Center for supporting him while he has been working on this study.